\documentclass[aps,prl,twocolumn,superscriptaddress,longbibliography]{revtex4-1}

\usepackage{pdfpages}

\makeatletter
\AtBeginDocument{\let\LS@rot\@undefined}
\makeatother

\usepackage{amsmath,amssymb,amsfonts}
\usepackage{graphicx}

\newcommand{\abs}[1]{\lvert #1 \rvert}
\newcommand{\expect}[1]{\langle #1\rangle}

\begin{document}
\title{Fluctuation-driven Coulomb drag in interacting quantum dot systems}
\author{Miguel A. Sierra}
\email{msierra@ifisc.uib-csic.es}
\author{David S\'anchez}
\affiliation{Instituto de F\'{\i}sica Interdisciplinar y Sistemas Complejos
  IFISC (UIB-CSIC), E-07122 Palma de Mallorca, Spain}
\author{Antti-Pekka Jauho}
\author{Kristen Kaasbjerg}
\email{kkaa@dtu.dk}
\affiliation{Center for Nanostructured Graphene (CNG), Department of Physics, Technical University of Denmark, DK-2800 Kgs. Lyngby, Denmark}

\begin{abstract}
  Coulomb drag between nanoscale conductors is of both fundamental and practical interest. Here, we theoretically study drag in a double quantum-dot (QD) system consisting of a biased drive QD and an unbiased drag QD coupled via a direct interdot Coulomb interaction. We demonstrate that the Coulomb drag is driven by charge fluctuations in the drive QD, and show how the properties of the associated quantum noise allow to distinguish it from, e.g., shot-noise driven drag in circuits of weakly interacting quantum conductors. In the strong-interaction regime exhibiting an orbital ("pseudospin") Kondo effect, the drag is governed by charge fluctuations induced by pseudospin-flip cotunneling processes. The quenching of pseudospin-flip processes by Kondo correlations are found to suppress the drag at low bias and introduce a zero-bias anomaly in the second-order differential transconductance. Finally, we show that the drag is maximized for values of the interdot interaction matching the lead couplings. Our findings are relevant for the understanding of drag in QD systems and provide experimentally testable predictions in different transport regimes.
\end{abstract}

\maketitle

\textbf{\emph{Introduction.}}---In recent years, systems of closely-spaced quantum-dots (QDs) or nanoscale conductors have been demonstrated to be hosts of novel transport mechanisms which can be exploited in, e.g., quantum information~\cite{Burkard1999}, thermoelectrics~\cite{Harman2002,Koski2015,Thierschmann2016}, and energy harvesting~\cite{Thierschmann2015}. Phenomena of fundamental importance such as, e.g., orbital Kondo physics~\cite{Borda2003,Jarillo-Herrero2005,Amasha2013} and attractive electron-electron interactions~\cite{Ilani2016} have been demonstrated. For such effects, the Coulomb interaction between the constituents is essential, and its appreciable size in nanoscale systems has driven experiments into hitherto inaccessible regimes.

This development has led to a revival of the phenomenon of Coulomb drag~\cite{Narozhny2016} in nanoscale systems with several reports of drag currents---i.e., a current induced in an unbiased drag system by its Coulomb interaction with a biased current-carrying drive system---in Coulomb-coupled double quantum dot (DQD) systems in 2DEGs~\cite{Shinkai2009,Keller2016} as well as in graphene and carbon nanotubes~\cite{Bischoff2015,Volk2015}. Theoretically, drag in quantum conductors~\cite{Mortensen2001,Mortensen2002,Levchenko2008,Borin2019} and QD systems~\cite{Sanchez2010,Kaasbjerg2016,Keller2016,Lim2018} has been studied thoroughly, with indications of an intimate link~\cite{Aguado2000,Levchenko2008} between drag and the quantum noise of nonequilibrium fluctuations~\cite{NazarovBook,Clerk2010}.

So far, theoretical works on drag in QD systems have been based on master-equation approaches in the Coulomb-blockade regime $\Gamma<k_B T, U$~\cite{Sanchez2010,Kaasbjerg2016,Keller2016,Lim2018}, which do not apply to quantum coherent transport and provide no direct interpretation in terms of quantum noise. In addition, experiments have demonstrated Coulomb drag across different transport regimes~\cite{Bischoff2015,Keller2016}, why further theoretical investigations may advance our understanding of Coulomb drag~\cite{Bischoff2015,Keller2016} and related transport effects~\cite{Harman2002,Koski2015,Thierschmann2015}, as well as the link to quantum noise.
\begin{figure}[b]
  \centering
  \includegraphics[scale=0.29]{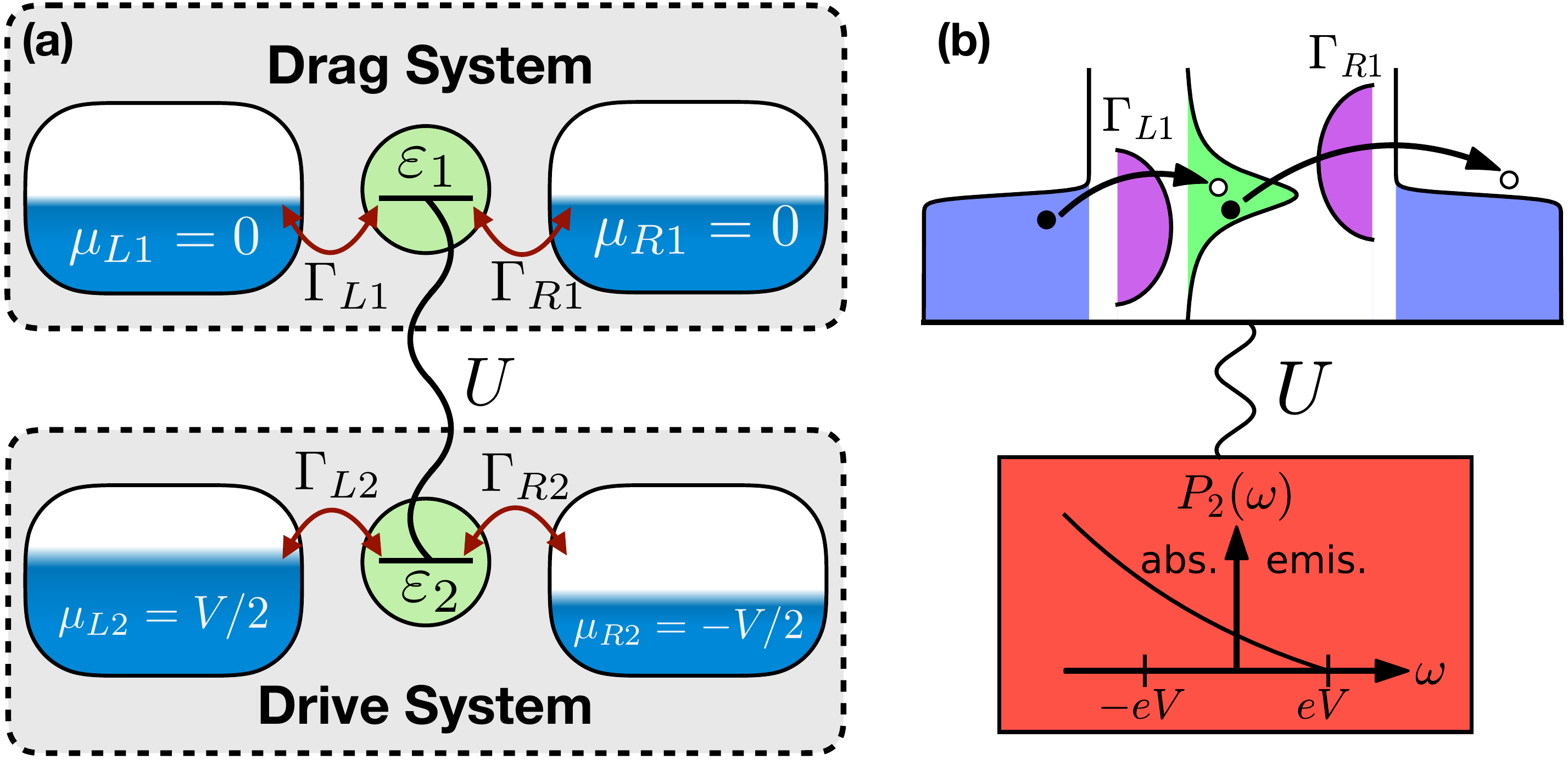}
  \caption{(a) Schematic illustration of the double quantum-dot setup consisting of an unbiased drag QD ($j=1$) and a biased drive QD ($j=2$) interacting via an interdot Coulomb interaction $U$  (no tunneling between the dots is allowed). (b) Energy diagram illustrating the charge-fluctuation driven drag in Eq.~\eqref{eq:Igtrless}. The quantum noise $P_2(\omega)$ of the charge fluctuations (with distinct absorption and emission components) in the biased drive system induces inelastic transitions in the drag system via the Coulomb interaction. This produces a drag current if the lead couplings $\Gamma_{L1}$ and $\Gamma_{R1}$ in the drag system have different energy dependencies.}
\label{fig:0}
\end{figure}

In this work, we apply the Keldysh nonequilibrium Green's function (NEGF) formalism~\cite{JauhoBook} to the description of Coulomb drag in QD systems across interaction regimes, covering weak ($\Gamma> U$), to intermediate ($\Gamma \sim U$), and strong ($\Gamma < U$) Coulomb interactions $U$, where $\Gamma$ is the overall lead coupling.

As we here demonstrate, the leading contribution to the drag is governed by the \emph{nonsymmetrized} quantum noise spectrum~\cite{NazarovBook} of the \emph{nonequilibrium} charge fluctuations in the biased drive system,
\begin{equation}\label{eq:Fluct}
  P_2(\omega) = \int dt\, e^{i\omega t} \expect{\bar{n}_2(0) \bar{n}_2(t) } ,
\end{equation}
where $\bar{n}_2 = \hat{n}_2 - \expect{\hat{n}_2}$ is the occupation of the drive dot relative to its mean value (see also Fig.~\ref{fig:0}). This is in stark contrast to the drag induced between two coherent conductors (e.g., QPCs) by a circuit environment~\cite{Levchenko2008}, which is driven by the quantum shot noise in the drive system; i.e., current fluctuations whose noise characteristics are distinctly different from those of charge fluctuations~\cite{Rothstein2008}. This points to a fundamental difference between drag mediated by, respectively, direct Coulomb interactions (this study) and a circuit environment (Ref.~\onlinecite{Levchenko2008}).

For strongly interacting DQD systems, higher-order processes involving tunneling events in both QDs become important leading to an orbital analog~\cite{Borda2003,Jarillo-Herrero2005,Amasha2013} of the conventional spin Kondo effect~\cite{Hewson,Ralph1994,vanderWiel2000,Rosch2001} where the QD levels play the role of a pseudospin. To lowest order in the effective exchange coupling we find that the drag is dominated by pseudospin-flip cotunneling processes corresponding to simultaneous charge fluctuations in both QD systems. The quenching of charge fluctuations in the Kondo regime is found to suppress the drag at low bias voltages, and leads to a zero-bias anomaly in the second-order derivative of the drag current with respect to the drive voltage.

\textbf{\emph{Model and theory.}}---We consider a spinless DQD system consisting of a drag ($j=1$) and drive ($j=2$) dot coupled via an interdot Coulomb interaction $U$ and connected to separate sets of source and drain contacts as depicted in Fig.~\ref{fig:0}. The total Hamiltonian takes the form
$\mathcal{H} = \mathcal{H}_\text{leads} + \mathcal{H}_\text{DQD} + \mathcal{H}_\text{tun}$,
where the Hamiltonian of the two Coulomb-coupled QDs is
$\mathcal{H}_\text{DQD} = \sum_j \varepsilon_{j} d_{j}^\dagger d_j + U
d_1^\dagger d_1 d_2^\dagger d_2$. Here, $\varepsilon_j$ is the
position of the gate-controlled energy level in the $j$th QD, and $U$ is the interdot Coulomb interaction. The contacts are described by noninteracting reservoirs, $\mathcal{H}_\text{leads} = \sum_{\alpha j k} \xi_{\alpha j k} c^\dagger_{\alpha j k} c_{\alpha j k}$, where $c^\dagger_{\alpha j k} $ ($c_{\alpha j k} $) creates (annihilates) an electron in state $k$ of lead $\alpha=\{L,R\}$ and system $j$ with energy $\xi_{\alpha j k}$. In the drive system, the chemical potentials of the reservoirs are given by the applied bias voltage, $\mu_\alpha \equiv \mu_{\alpha 2} = \varepsilon_F \pm eV_{\alpha}$, 
while the reservoirs of the drag system are kept in equilibrium with
$\mu_{\alpha 1}=\varepsilon_F=0$. Finally, the dot-reservoir tunneling is represented by $\mathcal{H}_{\rm tun} = \sum_{\alpha j k} \mathcal{V}_{\alpha j k} c^\dagger_{\alpha j k } d_{j}+ {\rm h.c.}$ where $\mathcal{V}_{\alpha j k}$ are the tunnel couplings.

We describe the QD system using NEGF where the contour-ordered dot Green's function (GF)
$G_j(\tau, \tau') = -i\expect{T_c {d_j(\tau)d_j^\dagger(\tau')}}$ is given by the usual Dyson equation with the irreducible self-energy, $\Sigma_{j} = \Sigma_{j,\mathrm{tun}} + \Sigma_{j,\mathrm{int}}$, having contributions from (i) the tunnel couplings to the leads, $\Sigma_{j,\mathrm{tun}}(\tau,\tau') = \sum_\alpha \Sigma_{\alpha j,\mathrm{tun}}(\tau,\tau') = \sum_{\alpha k} \left|\mathcal{V}_{\alpha j k}\right|^2 g_{\alpha j k}(\tau,\tau')$ where $g_{\alpha j k}$ is the unperturbed Green's function of the lead $\alpha j$, and (ii) the interdot Coulomb interaction, $\Sigma_{j,\mathrm{int}}=\Sigma_{j,H} + \Sigma_{j,\mathrm{xc}}$. The latter is split into separate Hartree ($H$) and exchange-correlation (xc) parts described in further detail below. Analytic continuation onto the real-time axis is performed with the Langreth rules~\cite{JauhoBook}.

The current in lead $\alpha j$ is defined as $I_{\alpha j} = -e \tfrac{d N_{\alpha j}}{dt}$, where $N_{\alpha j}$ is the total occupation of lead $\alpha j$, can be expressed in terms of the dot GF and self-energies in Fourier space as~\cite{JauhoBook}
\begin{align}
  \label{eq:CurrentSigmas}
  I_{\alpha j} & =
      \frac{e}{h}\int \! d\omega \, G^r_j(\omega)G^a_{j}(\omega) \nonumber \\
      & \quad 
      \times \left[
        \Sigma^>_{j}(\omega) \Sigma_{\alpha j,\mathrm{tun}}^<(\omega) - 
        \Sigma^<_{j}(\omega) \Sigma_{\alpha j,\mathrm{tun}}^>(\omega)
        \right] ,
\end{align}
where $G^{r/a}_{j}(\omega)=[\omega - \varepsilon_j - \Sigma_{j}^{r/a}(\omega)]^{-1}$
are the retarded/advanced dot Green's functions, and
$\Sigma_{\alpha j,\mathrm{tun}}^{r/a}(\omega) = \Lambda_{\alpha j}(\omega) \mp i \Gamma_{\alpha j}(\omega)/2$ are the retarded/advanced tunneling self-energies, $\Lambda_{\alpha j}$ gives the shift of the levels due to the tunnel coupling and $\Gamma_{\alpha j}$ is the lead hybridization function. The greater/lesser components are given by
$\Sigma_{\alpha j,\mathrm{tun}}^<(\omega) = - i \Gamma_{\alpha j}(\omega)
 f_{\alpha j}(\omega)$ and $\Sigma_{\alpha j,\mathrm{tun}}^>(\omega) = i \Gamma_{\alpha j}(\omega)
 [1- f_{\alpha j}(\omega)]$,
where $f_{\alpha j}(\omega)=1/[1+\exp{(\omega-\mu_{\alpha j})/k_B T}]$ is the
Fermi-Dirac distribution function.

From Eq.~\eqref{eq:CurrentSigmas} we note that the interacting self-energy is, not surprisingly, instrumental for a nonzero drag current~\cite{zerodrag}. More specifically, it is the dynamic xc part of the interaction self-energy which is responsible for the drag since the Hartree part has $\Sigma_{j, H}^{>/<}=0$~\cite{Thygesen2008}. On the other hand, the retarded/advanced Hartree self-energy, $\Sigma_{j,H}^{r,a}=U n_{\bar{j}}$, where
$n_j= -i \int \! \tfrac{d\omega}{2\pi} \, G^<_{j}(\omega)$ is the occupation of dot $j$, $\bar{j}\neq j$, and $G^<_{j}=G^r_{j}\Sigma^<_{j}G^a_{j}$, is essential for capturing the Coulomb coupling between the dots as it introduces an interaction-induced shift of the QD levels, $\varepsilon_j \rightarrow \varepsilon_j + U n_{\bar{j}}$.
\begin{figure}[t!]
  \centering
    \includegraphics[scale=0.39]{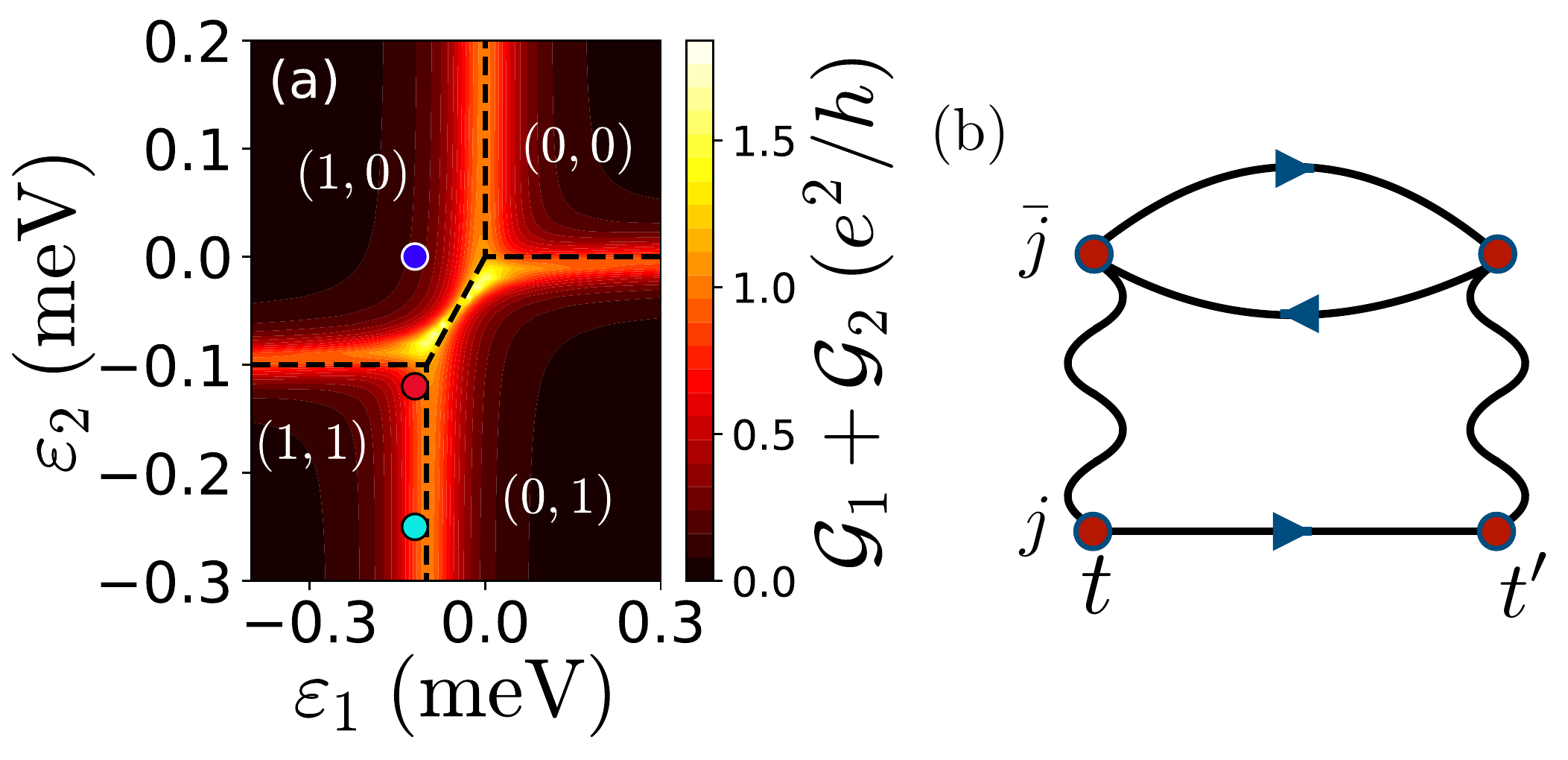}
  \caption{(a) Stability diagram showing the energetically most favorable dot
    occupations $(n_1,n_2)$ vs level positions (gate voltage). The honeycomb vertex (dashed lines) connecting the two triple points at the center of the diagram is due to the interdot
    Coulomb interaction $U$. The color scale shows the sum of the linear conductances $\mathcal{G}_1+\mathcal{G}_2$ in the drive and drag systems. The dots mark the level configurations considered in Fig.~\ref{fig:2}. Parameters: $U= 0.1 \;\mathrm{meV}$,
    $\gamma_{L1} = \gamma_{R1}= 7 \;\mu\mathrm{eV}$, $\gamma_{L2} = \gamma_{R2}= 25 \;\mu\mathrm{eV}$,
    $D_{\alpha j} = 10 \;\mathrm{meV}$, $\tilde{\varepsilon}_{L1}=-\tilde{\varepsilon}_{R1}= -2.5 \; \mathrm{meV}$ and $T=23 \;\mathrm{mK}$. (b) Feynman diagram for the single-bubble approximation to the xc part of interaction self-energy in Eq.~\eqref{eq:SigmaU2}.}
\label{fig:1}
\end{figure}

The effect of the Coulomb coupling is evident from the stability diagram in Fig.~\ref{fig:1}(a), which shows the sum of the linear conductances $\mathcal{G}_1+\mathcal{G}_2$, $\mathcal{G}_j=dI_j/dV_j$, as a function of the level positions $\varepsilon_j$ for a DQD system in the regime $\Gamma \sim U$. As indicated in Fig.~\ref{fig:1}(a), the stability diagram maps out the energetically most favorable occupations of the QDs, and is characterized by the Coulomb-induced honeycomb vertex (dashed lines) connecting the triple points at $\varepsilon_j=-U$ and $\varepsilon_j=0$~\cite{Wiel2002}.

For the xc part of the interaction self-energy, we adopt the single-bubble approximation illustrated by the Feynman diagram in Fig.~\ref{fig:1}(b), and given by
\begin{equation}
  \label{eq:SigmaU2}
  \Sigma_{j , \text{xc}}(\tau, \tau') = U^2 G_j(\tau, \tau') P_{\bar{j}}(\tau, \tau'),
    \quad \bar{j} \neq j,
\end{equation}
where $P_j(\tau, \tau')=G_j(\tau, \tau')G_j(\tau', \tau)$ is
the nonequilibrium polarization bubble of the $j$ system. The greater and lesser components of the self-energy are $\Sigma_{j , \text{xc}}^{>/<}(t,t') = U^2 G_j^{>/<}(t,t') P_{\bar{j}}^{>/<}(t,t')$, where $P_{j}^{>/<}(t,t') = G_{j}^{>/<}(t,t')G_{j}^{</>}(t',t)$. Generalization to more complicated self-energies, such as, e.g., the $GW$ approximation~\cite{Thygesen2008}, is possible~\cite{Tanatar2009,Zhou2019}. However, the $GW$ approximation is more relevant for extended systems where screening effects are important, and is therefore not expected to affect our findings below.

In the following, we pursue both analytic perturbative and numerical nonperturbative calculations~\cite{Lunde2009}. In the numerical calculations, the xc self-energy in Eq.~\eqref{eq:SigmaU2} is obtained from the Hartree GF. Since the latter itself depends on the dot occupations, the Hartree GF must be calculated self-consistently. We should stress that this procedure yields a conserving approximation with overall charge conservation respected~\cite{supplemental}, $\sum_\alpha I_{\alpha j}=0$, allowing us to define the drive and drag currents as $I_{j}\equiv(I_{Lj}-I_{Rj})/2$. 

\textbf{\emph{The drag current.}}---We start by analyzing the drag arising from the self-energy in Eq.~\eqref{eq:SigmaU2}. Inserting in the general expression for the current in Eq.~\eqref{eq:CurrentSigmas}, the drag current can be written as~\cite{supplemental}
\begin{align}
\label{eq:Igtrless}
    I_\text{drag} & = \frac{eU^2}{h} \int \! d\omega \frac{d\omega'}{2\pi} \, 
      \frac{\mathcal{A}_1(\omega)\mathcal{A}_1(\omega-\omega')}
                  {\Gamma_1(\omega)\Gamma_1(\omega-\omega')}
    \nonumber \\ 
    & \quad\times 
        \left[\Gamma_{L 1}(\omega-\omega') \Gamma_{R 1}(\omega)-
              \Gamma_{L 1}(\omega) \Gamma_{R 1}(\omega-\omega') 
      \right]
      \nonumber \\
      & \quad \times
      f_1 (\omega-\omega') \left[ 1 - f_1(\omega)\right] P_{2}^<(\omega')      \; ,
\end{align}
where $\mathcal{A}_j(\omega) = -(1/\pi) \mathrm{Im}\, G^r_{j}(\omega)$ is the dot spectral function and $\Gamma_j = \Gamma_{Lj}+\Gamma_{Rj}$.
\begin{figure}[t!]
  \centering
  \includegraphics[width=\linewidth]{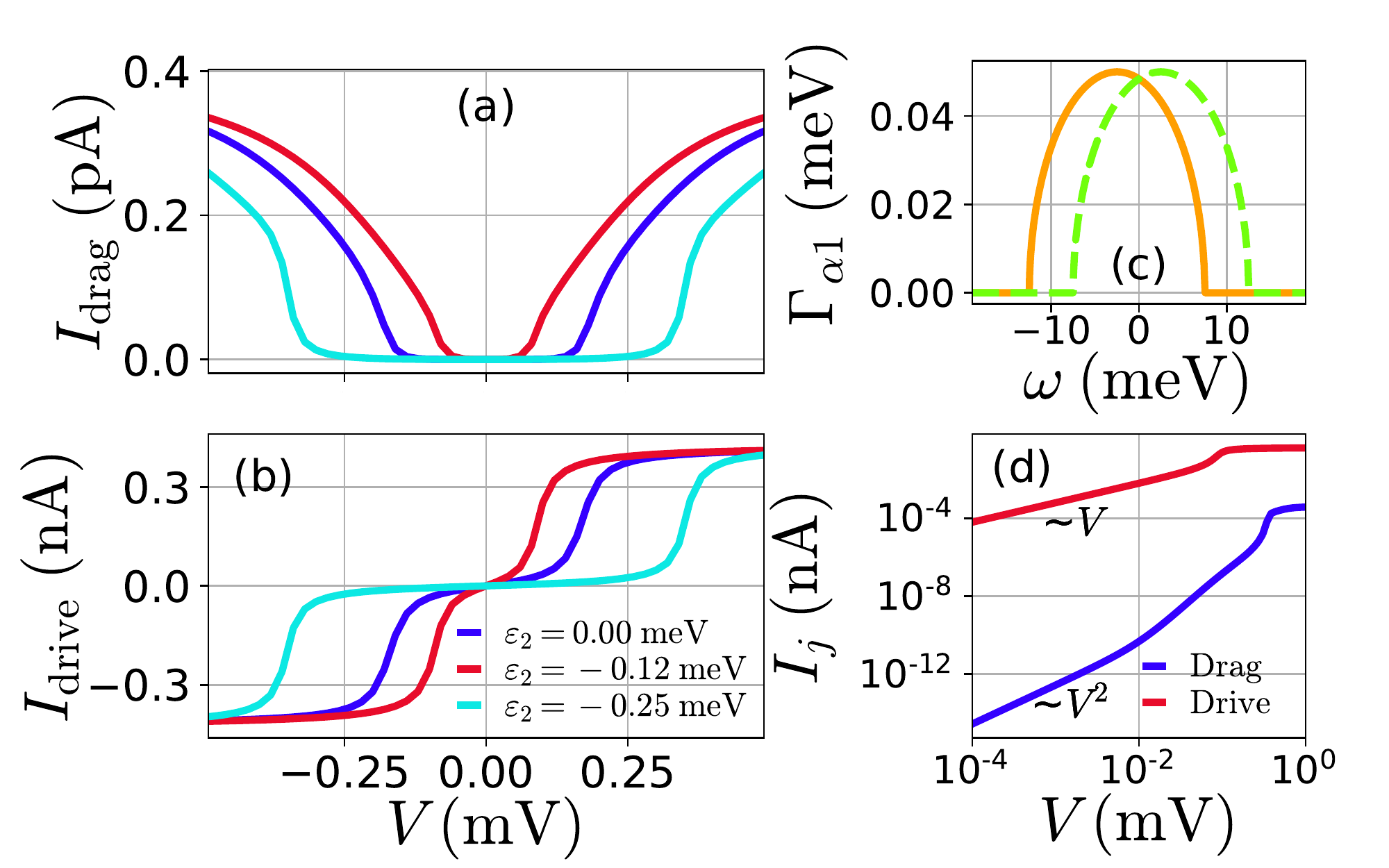}
  \caption{(a) Drag current, and (b) drive current as a function of the applied bias voltage in the drive system for the level positions indicated with dots in Fig.~\ref{fig:1}(a). (c) Lead hybridization functions $\Gamma_{\alpha j}$ modelled by semielliptic bands. As illustrated, the bands in the left (full) and right (dashed) leads of the drag system are shifted relative to each other in order to yield a nonzero drag. (d) Log-log plot of the drag and drive currents vs bias voltage showing the leading contributions of $V$. Same parameters as in Fig.~\ref{fig:1} and $\varepsilon_{1}=-0.12 \;\mathrm{meV}$.}
\label{fig:2}
\end{figure}

Several important observations can be made from Eq.~\eqref{eq:Igtrless}. First, in order for a nonzero drag current, the lead couplings in the drag QD must fulfill $\Gamma_{L1}(\omega) \Gamma_{R1}(\omega' ) - \Gamma_{L1}(\omega') \Gamma_{R1}(\omega) \neq 0$; i.e., they cannot be proportional $\Gamma_{L1}(\omega)\neq C \Gamma_{R1}(\omega)$. This result is consistent with previous works in the Coulomb blockade regime~\cite{Kaasbjerg2016,Keller2016}. Secondly, the drag can be understood as arising from interaction-mediated creation and annihilation of electron-hole pair excitations in the drag and drive systems, and is driven by the finite-bias correlator $P_2^<$ of the drive QD which can be identified as the quantum charge noise $P_2(\omega)$ in Eq.~\eqref{eq:Fluct}.
At low temperature, i.e., $k_BT \ll eV$, where the drag QD predominantly absorbs energy from the drive QD due to Pauli blocking of emission processes, the quantum nature of the noise manifests itself in the fact that the drag is governed by the \emph{emission noise} of the drive QD~\cite{Gavish2000,NazarovBook} given by $P_2(\omega > 0)$~\cite{supplemental}. Lastly, the drag in Eq.~\eqref{eq:Igtrless} can be viewed as a rectification of the quantum charge noise in the drive system, which discerns it from the shot-noise driven drag of Ref.~\cite{Levchenko2008} mentioned in the introduction.

\textbf{\emph{Weak-interaction regime.}}---Having established the general properties of the drag, we next examine its behavior for weak interactions ($U\lesssim \Gamma$).

In the small-$U$ limit, $U \ll \Gamma$, an explicit expansion of the drag current to second order in $U$ applies, and amounts to replacing $G_j\rightarrow G_{j}^0$, where $G_{j}^0$ is the \emph{noninteracting} dot GF, in Eq.~\eqref{eq:Igtrless}. In this limit, our DQD system behaves similarly as two Coulomb-coupled single-channel QPCs with transmission coefficients $T_j= 4\Gamma_{jL}\Gamma_{jR} / \Gamma_j^2$, and the charge noise scales as $P_2 \sim  T_2$ for $k_B T \ll eV\ll \Gamma$~\cite{supplemental}. This should be contrasted with the qualitatively different $S_2\sim T_2 (1-T_2)$ scaling of the shot noise~\cite{Blanter2000}, thus providing a means to distinguish between drag due to, respectively, direct Coulomb interactions and a circuit environment~\cite{Levchenko2008} by tuning the conductance $\mathcal{G}_2=\tfrac{e^2}{h} T_2$ of the drive system.

Next, we turn to numerical calculations of the drag using parameters mimicking the experiment in Ref.~\cite{Keller2016}. In order to fulfil the conditions for a nonzero drag, we model for convenience the lead couplings $\Gamma_{\alpha j}$ by semielliptic bands with bandwidth $D_{\alpha j}$ and a relative shift between the bands in the left and right contacts of the drag system as illustrated in Fig.~\ref{fig:2}(c)~\cite{supplemental}. 

Figure.~\ref{fig:2}(a) and~\ref{fig:2}(b) show the calculated drag and drive currents for the level positions marked with the colored dots in Fig.~\ref{fig:1}(a). At low bias, the level of the drive dot is off resonance with respect to the chemical potentials and both the drive and drag currents are small. With increasing bias voltage, the level of the drive dot enters the conduction window and the onset of the drive current induces a current in the drag system. While the direction of the drive current follows the sign of the applied bias voltage, the direction of the drag is governed by the lead couplings. From Eq.~\eqref{eq:Igtrless}, this can be traced back to the fact that $I_{\mathrm{drag}}$ has no linear dependence on $V$ when the drive contacts have proportional lead couplings, $\Gamma_{L2} \propto \Gamma_{R2}$~\cite{supplemental}. As demonstrated in Fig.~\ref{fig:2}(d), the drag current at low bias thus increases as $I_{\mathrm{drag}}\sim V^2$, emphasizing the inherent nonlinear nature of Coulomb drag in quantum dot systems~\cite{Kaasbjerg2016}.

\begin{figure}[t!]
  \centering
  \includegraphics[scale=0.33]{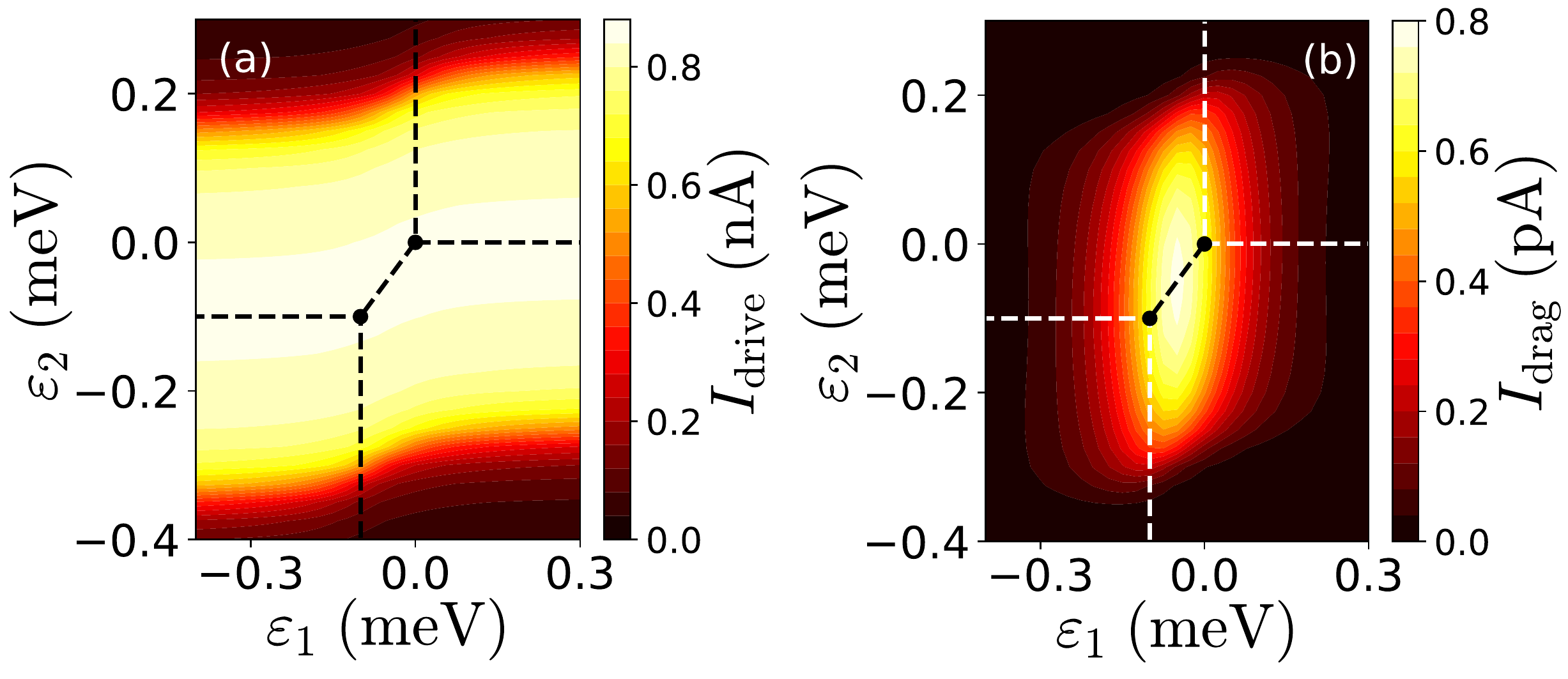}
  \caption{Drive (a) and drag (b) currents as a function of the level positions $\varepsilon_1$ and $\varepsilon_2$ for an applied bias of $V=0.5$~mV. All other parameters are the same as in Fig.~\ref{fig:1}.}
\label{fig:3}
\end{figure}
The dependence of the currents on the QD levels at a fixed bias voltage of $V=0.5 \,\mathrm{mV}$ is shown in Fig.~\ref{fig:3}. Here, the drive current in Fig.~\ref{fig:3}(a) shows a weak dependence on the drag level, $\varepsilon_{1}$, due to the interaction-induced shift of the drive level upon changing the occupation of the drag QD. The drag current in Fig.~\ref{fig:3}(b) is only significant near the honeycomb vertex where the alignment between the QD levels and the chemical potentials allows for simultaneous interaction-induced electron-hole pair processes in the drive and drag system [cf. Fig.~1(b)]. Our results in Figs.~\ref{fig:1}(a) and~\ref{fig:3} are in good qualitative agreement with the experimental measurements reported in Ref.~\cite{Keller2016} [their Figs.~2(b)-(c) and 2(e)-(f)].

\textbf{\emph{Strong-interaction and Kondo regime.}}---In the regime of strong interdot interaction, $U\gg \Gamma$, the DQD Hamiltonian can be mapped onto the Kondo Hamiltonian $\mathcal{H}_{\rm K}=\mathcal{H}_{\rm dir} + \mathcal{H}_{\rm ex}$ by a Schrieffer-Wolff transformation~\cite{Schrieffer1966,supplemental} with the two QD states acting as a pseudospin ($\hat{S}$) exhibiting an orbital Kondo effect~\cite{Borda2003,Jarillo-Herrero2005,Amasha2013} at $T<T_K$ where $T_K \sim \sqrt{\Gamma U } \exp(-\pi U/4\Gamma )$ is the Kondo temperature \cite{Kaminski2000}. 
Here, $\mathcal{H}_{\rm dir}=\sum_{\alpha \beta j kq } \mathcal{K}_{\alpha j,\beta j} c^\dagger_{\alpha j k} c_{\beta j q}^{\phantom\dagger}$ is the 
potential scattering term responsible for \emph{elastic} pseudospin-conserving cotunneling transitions, whereas the exchange interaction $\mathcal{H}_{\rm ex}=\sum_{\alpha \beta ij kq }\mathcal{J}_{\alpha i, \beta j} \hat{S}_l s^l_{ij} c^\dagger_{\alpha i k} c_{\beta j q}^{\phantom\dagger}$ accounts for, e.g.,  \emph{inelastic} pseudospin-flip cotunneling processes, and $\mathcal{K}_{\alpha i,\beta j}$ and $\mathcal{J}_{\alpha i, \beta j}$ are the effective couplings~\cite{Hewson,supplemental}.

The importance of pseudospin (i.e., charge) fluctuations inherent to the orbital Kondo effect is evident already from the leading-order $I_\text{drag}\propto \abs{\mathcal{J}}^2$ contribution to the drag. Performing a perturbative calculation of the drag at the particle-hole symmetric point ($\varepsilon_1=\varepsilon_2=-U/2$)~\cite{Kaminski2000}, we find~\cite{supplemental},
\begin{align}
  \label{eq:DragPert}
  I_\text{drag} & = 
      \frac{e}{2h}\sum_\beta \int\! d\omega \, \Delta \Gamma_1 \Gamma_{\beta 2} \left|\frac{1}{\omega+U/2}-\frac{1}{\omega-U/2}\right|^2
      \nonumber \\
  & \quad \times [f_1(\omega)-f_{\beta 2}(\omega)] ,
\end{align}
where $\Delta\Gamma_j=\Gamma_{Lj}-\Gamma_{Rj}$. This can be viewed as drag due to \emph{nonlocal} cotunneling processes where the pseudospin of the DQD is flipped in one coherent processes via virtual intermediate empty and filled states~\cite{Kaasbjerg2016,Lim2018}. 
\begin{figure}[!t]
  \centering
  \includegraphics[scale=0.45]{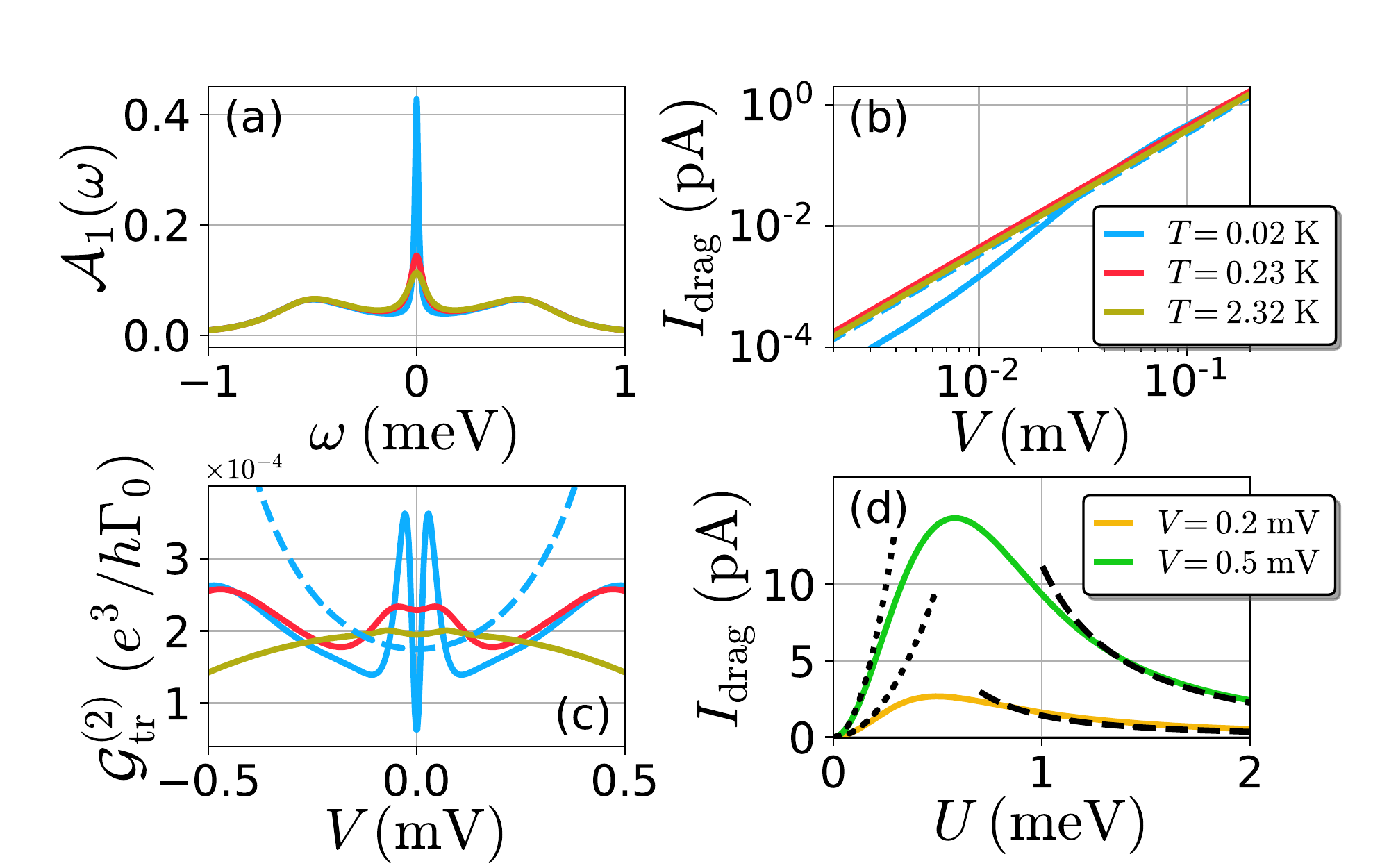}
  \caption{(a) Spectral function of the drag system at zero drive bias, (b) drag current vs drive bias, and (c) second-order differential transconductance $\mathcal{G}_{\rm tr}^{(2)}=d^2I_{\rm drag}/dV^2$ of the drag current vs drive bias, all at the particle-hole symmetric point and for the temperatures indicated in (b). (d) Drag current as a function of the interdot interaction for two different bias voltages. Solid lines correspond to numerical calculations following Eq.~\eqref{eq:Igtrless}, while dotted and dashed lines show, respectively, the leading-order contribution to Eq.~\eqref{eq:Igtrless} and the perturbative result in Eq.~\eqref{eq:DragPert}. Parameters: $U = 1 \; {\rm meV}$, $\varepsilon_1=\varepsilon_2=-U/2$, $\gamma_{\alpha j} = 0.05 \; {\rm meV}$, $\tilde{\varepsilon}_{L1} = -\tilde{\varepsilon}_{R1} = -4\; {\rm meV}$, $\tilde{\varepsilon}_{L2}=\tilde{\varepsilon}_{R2}=0$ and $D_{\alpha j} = 10 \; {\rm meV}$.}
\label{fig:4}
\end{figure}

To get an indication of the corrections to the drag in Eq.~\eqref{eq:DragPert} due to Kondo correlations, we show in Fig.~\ref{fig:4} numerical results [based on Eq.~\eqref{eq:Igtrless}] for the drag at different temperatures at the particle-hole symmetric point. The spectral function $\mathcal{A}_1(\omega)$ of the drag dot in Fig.~\ref{fig:4}(a) clearly shows a low-temperature feature at the Fermi energy which resembles a Kondo peak~\cite{Horvatic1987,Sun2002,Bao2014}. Overall, the corresponding drag currents in Fig.~\ref{fig:4}(b) only show marginal changes relative to the perturbative result in Eq.~\eqref{eq:DragPert} (dashed line). However, at low temperature and low bias, Kondo correlations quench the charge fluctuations driving the drag which results in a suppression relative to the perturbative result. Due to the low-bias $I_\mathrm{drag}\sim V^2$ behavior of the drag, the effect of Kondo correlations is more noticeable in the second-order differential transconductance $\mathcal{G}_{\rm tr}^{(2)}=d^2I_{\rm drag}/dV^2$ shown in Fig.~\ref{fig:4}(c), which features a pronounced zero-bias anomaly in $\mathcal{G}_{\rm tr}^{(2)}$ at low $T$. Finally, Fig.~\ref{fig:4}(d) shows the dependence of the drag on the strength of the interdot interaction at different bias voltages. For large values of $U$, the numerical results follow the $I_{\rm drag}\sim 1/U^2$ behavior of the perturbative result of Eq.~\eqref{eq:DragPert} (dashed lines), whereas the $I_{\rm drag}\sim U^2$ behavior of the leading-order contribution to the drag in Eq.~\eqref{eq:SigmaU2} is observed for small values of $U$ (dotted line). The two regimes are bridged by an intermediate region, $\Gamma\sim U$, with an optimal value of $U$ where $|I_{\rm drag}|$ is maximized.

\textbf{\emph{Conclusions.}}---We have studied Coulomb drag across interaction regimes in Coulomb-coupled QD systems in the framework of the Keldysh NEGF technique. In agreement with previous works~\cite{Sanchez2010,Kaasbjerg2016,Keller2016,Lim2018}, we find that drag is an inherently nonlinear effect and that energy-dependent lead couplings are instrumental for the generation of a drag current. As we demonstrate, the drag is driven by the nonequilibrium charge fluctuations of the drive QD, and we discuss how the characteristics of the quantum noise allows to differentiate the drag mechanism discussed here from drag induced by a circuit environment~\cite{Levchenko2008} experimentally. In the case of strong interdot interactions, the charge fluctuations are quenched by orbital Kondo correlations at low temperature and bias, which suppresses the drag current with respect to the lowest-order cotunneling-only drag~\cite{Kaasbjerg2016}. In addition, we predict a clear signature of Kondo correlations in the second-order differential transconductance of the drag.
Overall, our findings show that the Coulomb drag between quantum conductors is highly dependent on the interaction mediating the drag, and open the opportunity for further experimental studies of drag in QD systems.

\begin{acknowledgments}
  \textbf{\emph{Acknowledgements}}---We would like to thank J.~Paaske for fruitful discussions. The work was supported by the CAIB predoctoral grant and the fellowships for research internships of the government of Balearic Islands. D.S. acknowedges support from grant MAT2017-82639. K.K. acknowledges support from the European Union's Horizon 2020 research and innovation programme under the Marie Sklodowska-Curie Grant Agreement No.~713683 (COFUNDfellowsDTU). The Center for Nanostructured Graphene (CNG) is sponsored by the Danish National Research Foundation, Project DNRF103.
\end{acknowledgments}

\bibliographystyle{apsrev}
\bibliography{total}

\begin{thebibliography}{44}
\bibitem [{zer()}]{zerodrag}%
  \BibitemOpen
  \href@noop {} {}\bibinfo {note} {The drag current~\eqref{eq:CurrentSigmas} is
  zero for $\Sigma_{j,\mathrm{int}}=0$, which follows straight-forwardly from
  the identity $\Sigma^>_{Lt,\mathrm{tun}} \Sigma_{Rt,\mathrm{tun}}^< =
  \Sigma^<_{Lt,\mathrm{tun}}\Sigma_{Rt,\mathrm{tun}}^>$ for the tunnel
  self-energies in the unbiased drag system.}\BibitemShut {Stop}%
\end{thebibliography}

\pagebreak
\widetext
\clearpage


\section*{Supplementary material}

# Supplemental material: Fluctuation-driven Coulomb drag in interacting quantum dot systems

Miguel A. Sierra,[1] David Sánchez,[1] Antti-Pekka Jauho,[2] and Kristen Kaasbjerg[2]

[1]*Instituto de Física Interdisciplinar y Sistemas Complejos IFISC (UIB-CSIC), E-07122 Palma de Mallorca, Spain*
[2]*Center for Nanostructured Graphene (CNG), Department of Physics,*
*Technical University of Denmark, DK-2800 Kgs. Lyngby, Denmark*

## I. CHARGE CONSERVATION

In general, non-selfconsistent calculations of the interacting self-energy $\Sigma_{j,\text{int}}$ may lead to expressions for the current which do not fulfill overall charge conservation. In our work, charge conservation can be expressed as $I_{Lj} + I_{Rj} = 0$ for the separate drive and drag systems, $j = 1, 2$. In this section we show that our approach respects charge conservation in spite of the fact that the self-energy is not calculated fully self-consistently.

Performing the sum over leads and inserting the full self-energy $\Sigma_j = \Sigma_{j,\text{tun}} + \Sigma_{j,\text{int}}$ into the current expression given in Eq. (2) of the main text, we obtain

$$\sum_\alpha I_{\alpha j} = \frac{e}{h} \int d\omega \, G_j^r(\omega) G_j^a(\omega) \left[ \Sigma_{j,\text{int}}^>(\omega) \Sigma_{j,\text{tun}}^<(\omega) - \Sigma_{j,\text{int}}^<(\omega) \Sigma_{j,\text{tun}}^>(\omega) \right]. \tag{S1}$$

Next, inserting the interaction self-energy $\Sigma_{j,\text{int}} = \Sigma_{j,H} + \Sigma_{j,\text{xc}}$ with $\Sigma_{j,\text{xc}}$ approximated by the single-bubble self-energy in Eq. (3), we obtain

$$\sum_\alpha I_{\alpha j} = \sum_\alpha (I_{\alpha j}^> + I_{\alpha j}^<) = \frac{eU^2}{4\pi^2 h} \int d\omega d\omega_1 d\omega_2 \, |G_j^r(\omega)|^2 \left[ G_j^>(\omega - \omega_1) G_j^>(\omega_1 - \omega_2) G_j^<(-\omega_2) \Sigma_{j,\text{tun}}^<(\omega) \right]$$
$$- \frac{eU^2}{4\pi^2 h} \int d\omega d\omega_1 d\omega_2 \, |G_j^r(\omega)|^2 \left[ G_j^<(\omega - \omega_1) G_j^<(\omega_1 - \omega_2) G_j^>(-\omega_2) \Sigma_{j,\text{tun}}^>(\omega) \right]. \tag{S2}$$

In this expression, the GFs in the square brackets originate from the lesser/greater self-energies in Eq. (1). Since the self-energies in our non-selfconsistent approach are obtained from the Hartree GF, the lesser and greater GFs in Eq. (S2) can hence be replaced by $G_j^{<,>}(\omega) = G_j^r(\omega) \Sigma_{j,\text{tun}}^{<,>}(\omega) G_j^a(\omega)$ which has no contribution from the Hartree self-energy as $\Sigma_{j,H}^{>/<} = 0$.

With this, Eq. (S2) becomes

$$\sum_\alpha I_{\alpha j}^> = \frac{eU^2 4\pi^2}{h} \int d\vec{\omega} \, \tilde{\mathcal{A}}_{j,\bar{j}}(\omega, \omega_1, \omega_2) \left[ \Sigma_{j,\text{tun}}^>(\omega - \omega_1) \Sigma_{j,\text{tun}}^>(\omega_1 - \omega_2) \Sigma_{j,\text{tun}}^<(-\omega_2) \Sigma_{j,\text{tun}}^<(\omega) \right] \tag{S3}$$

$$\sum_\alpha I_{\alpha j}^< = -\frac{eU^2 4\pi^2}{h} \int d\vec{\omega} \, \tilde{\mathcal{A}}_{j,\bar{j}}(\omega, \omega_1, \omega_2) \left[ \Sigma_{j,\text{tun}}^<(\omega - \omega_1) \Sigma_{j,\text{tun}}^<(\omega_1 - \omega_2) \Sigma_{j,\text{tun}}^>(-\omega_2) \Sigma_{j,\text{tun}}^>(\omega) \right], \tag{S4}$$

where $d\vec{\omega} = d\omega d\omega_1 d\omega_2$ and

$$\tilde{\mathcal{A}}_{j,\bar{j}}(\omega, \omega_1, \omega_2) = \frac{\mathcal{A}_j(\omega) \mathcal{A}_j(\omega - \omega_1) \mathcal{A}_{\bar{j}}(\omega_1 - \omega_2) \mathcal{A}_{\bar{j}}(-\omega_2)}{\Gamma_j(\omega) \Gamma_j(\omega - \omega_1) \Gamma_{\bar{j}}(\omega_1 - \omega_2) \Gamma_{\bar{j}}(-\omega_2)}. \tag{S5}$$

The only difference between the two current contributions in Eqs. (S3) and (S4) is the energy dependence of the self-energies and a overall sign. With an appropriate change of the integration variables ($\omega' = \omega - \omega_1$, $\omega_1' = -\omega_1$ and $\omega_2' = \omega_2 - \omega_1$) we find that $I_{\alpha j}^> = -I_{\alpha j}^<$, implying that $\sum_\alpha I_{\alpha j} = 0$. Therefore, this approach respects charge conservation.

## II. EXPRESSION FOR THE DRAG CURRENT

In this section we show the algebraic manipulations which lead to the final expression for the current in Eq. (4) of the main text. First, we consider the lesser and greater components of the current defined above and express them in terms of the lesser and greater components of the polarization bubble,

$$I_{\alpha j}^{>,<} = \pm \frac{eU^2}{2\pi h} \int d\omega d\omega_1 \frac{\mathcal{A}_j(\omega) \mathcal{A}_j(\omega - \omega_1)}{\Gamma_j(\omega) \Gamma_j(\omega - \omega_1)} \left[ \Sigma_{\alpha j,\text{tun}}^{<,>}(\omega) \Sigma_{j,\text{tun}}^{>,<}(\omega - \omega_1) P_{\bar{j}}^{>,<}(\omega_1) \right], \tag{S6}$$

The tunnel self-energies are given by

$$\Sigma^<_{\alpha j,\text{tun}}(\omega) = i\Gamma_{\alpha j}(\omega)f_{\alpha j}(\omega) \quad \text{and} \quad \Sigma^>_{\alpha j,\text{tun}}(\omega) = -i\Gamma_{\alpha j}(\omega)[1 - f_{\alpha j}(\omega)],$$ (S7)

and hence Eq. (S6) becomes

$$I^{>,<}_{\alpha j} = \pm\frac{eU^2}{2\pi h}\sum_{\beta}\int d\omega d\omega_1 \frac{\mathcal{A}_j(\omega)\mathcal{A}_j(\omega - \omega_1)}{\Gamma_j(\omega)\Gamma_j(\omega - \omega_1)}\Gamma_{\alpha j}(\omega)\Gamma_{\beta j}(\omega - \omega_1)\left[\mathcal{F}^{>,<}_{\alpha\beta j}(\omega, \omega - \omega_1)P^{>,<}_j(\omega_1)\right],$$ (S8)

where $\mathcal{F}^>_{\alpha\beta j}(\omega, \omega') = f_{\alpha j}(\omega)\bar{f}_{\beta j}(\omega')$, $\mathcal{F}^<_{\alpha\beta j}(\omega, \omega') = \bar{f}_{\alpha j}(\omega)f_{\beta j}(\omega')$ and $\bar{f}_{\alpha j} = 1 - f_{\alpha j}$. The expression for the current can be obtained by performing a change of variables in the greater component of the drag current ($\omega' = \omega - \omega_1$ and $\omega'_1 = -\omega_1$) by which we obtain

$$I^>_{\alpha 1} = \frac{eU^2}{2\pi h}\int d\omega' d\omega'_1 \frac{\mathcal{A}_1(\omega')\mathcal{A}_1(\omega' - \omega'_1)}{\Gamma_1(\omega')\Gamma_1(\omega' - \omega'_1)}\Gamma_{\alpha 1}(\omega' - \omega'_1)\Gamma_{\beta 1}(\omega')f_1(\omega' - \omega'_1)\bar{f}_1(\omega')P^>_2(-\omega'_1).$$ (S9)

Taking into account the relation $P^>_2(-\omega) = P^<_2(\omega)$ between the lesser and greater components of the polarization bubble, the sum of the two current components becomes

$$I_{\alpha 1} = \frac{eU^2}{2\pi h}\int d\omega d\omega_1 \left[\Gamma_{\alpha 1}(\omega - \omega_1)\Gamma_{\beta 1}(\omega) - \Gamma_{\alpha 1}(\omega)\Gamma_{\beta 1}(\omega - \omega_1)\right]\frac{\mathcal{A}_1(\omega)\mathcal{A}_1(\omega - \omega_1)}{\Gamma_1(\omega)\Gamma_1(\omega - \omega_1)}f_1(\omega - \omega_1)\bar{f}_1(\omega)P^<_2(\omega_1),$$ (S10)

which coincides with Eq. (4) of the main text following the total expression of $I_{\text{drag}} = (I_{L1} - I_{R1})/2$.

From this expression, one notes that at $T = 0$ the integration range is restricted by the Fermi functions in the drag leads to $\omega > 0$ and $\omega_1 > \omega$. As a consequence, the drag depends only on $P^<_2(\omega > 0)$ which is the so-called emission part of a quantum noise spectrum. The physical interpretation of this observation is the following: at $T = 0$ the drag system ($j = 1$) can only *absorb* energy from the drive system ($j = 2$), and hence the drag current can only depend on the part of the quantum noise spectrum of the drive system which describes its *emission* properties.

### III. CHARGE NOISE IN THE WIDE-BAND LIMIT

In the limit where the lead couplings by far exceed any of the other energy scales in the problem, i.e., $\Gamma_j \gg k_B T, eV, U, \omega$, the *noninteracting* quantum noise for the charge fluctuations in system $j$ becomes

$$P^<_j(\omega) = \int\frac{d\omega'}{2\pi}G^<_j(\omega')G^>_j(\omega' - \omega) = \int\frac{d\omega'}{2\pi}\frac{\sum_\alpha \Gamma_{\alpha j}f_{\alpha j}(\omega')}{(\omega' - \varepsilon_j)^2 + (\Gamma_j/2)^2}\frac{\sum_\beta \Gamma_{\beta j}[1 - f_{\beta j}(\omega' - \omega)]}{(\omega' - \omega - \varepsilon_j)^2 + (\Gamma_j/2)^2}$$

$$\approx \frac{1}{(\Gamma_j/2)^4}\times\begin{cases} -\omega\sum_{\alpha\beta}\Gamma_{\alpha j}\Gamma_{\beta j}, & \omega < -e|V| \\ -\omega\sum_\alpha \Gamma^2_{\alpha j} + \Gamma_{Lj}\Gamma_{Rj}(e|V| - \omega), & -e|V| < \omega < 0 \\ \Gamma_{Lj}\Gamma_{Rj}(e|V| - \omega), & 0 < \omega < e|V| \\ 0, & e|V| < \omega \end{cases}.$$ (S11)

While this has the same $\omega$ dependence as the finite-frequency quantum shot noise [S1], the scaling with the transmission coefficient $T_j = 4\Gamma_{Lj}\Gamma_{Rj}/\Gamma_j^2$ is different. For the charge fluctuations the emission noise ($\omega > 0$) scales as $P^<_j \sim T_j/\Gamma_j^2$ while the shot noise has the well-known $S_j \sim T_j(1 - T_j)$ behavior.

### IV. NEWNS-ANDERSON MODEL

As explained in the main text, the double quantum dot system requires nonproportional hybridization functions in order to find nonvanishing drag currents. This is only satisfied if the contacts of the system are energy-dependent. For this reason, tight-binding contacts with shifted bandwidth has been considered in the main manuscript. Hence, the corresponding retarded tunneling self-energies are obtained employing the Newns-Anderson model:

$$\Sigma^r_{\alpha j,\text{tun}}(\omega) = \gamma_{\alpha j}\left[\tilde{\omega}_{\alpha j} + \text{sgn}(\tilde{\omega}_{\alpha j})\theta(|\tilde{\omega}_{\alpha j}| - 1)\sqrt{\tilde{\omega}^2_{\alpha j} - 1} - i\theta(1 - |\tilde{\omega}_{\alpha j}|)\sqrt{1 - \tilde{\omega}^2_{\alpha j}}\right],$$ (S12)

where $\tilde{\omega}_{\alpha j} = (\omega - \tilde{\varepsilon}_{\alpha j})/D_{\alpha j}$. Here, $D_{\alpha j}$, $\tilde{\varepsilon}_{\alpha j}$ and $\gamma_{\alpha j}$ are the bandwidth, the shift with respect to $\omega = \varepsilon_F = 0$ and the tunneling amplitude of the lead $\alpha = \{L, R\}$, $j = \{1, 2\}$. Thus, the nonproportionality in the drag system is achieved assuming different shifts $\varepsilon_{L1} \neq \varepsilon_{R1}$.

## V. SCHRIEFFER-WOLFF TRANSFORMATION

In the Kondo regime, we have compared our results with calculations of the current given by a perturbation expansion of the Kondo Hamiltonian. Both Anderson and Kondo models are related with each other via the Shrieffer-Wolff transformation [S2] which is a unitary transformation which discards sequential tunneling processes. In contrast with Ref. [S2], we extend the model allowing different levels $\varepsilon_{1,2}$ between quantum dots.

The procedure consists of a transformation of the Anderson Hamiltonian such as $\mathcal{H} \rightarrow \mathcal{H}' = e^{\mathcal{S}} \mathcal{H} e^{-\mathcal{S}}$ in which the operator $\mathcal{S}$ reads

$$\mathcal{S} = \sum_{\alpha j k} \left[ w^{(1)}_{\alpha j k} n_{\bar{j}} C^{\dagger}_{\alpha j k} d_j + w^{(2)}_{\alpha j k} (1 - n_{\bar{j}}) d^{\dagger}_j C_{\alpha j k} - \text{h.c.} \right] , \tag{S13}$$

where $w^{(1)}_{\alpha j k}$ and $w^{(2)}_{\alpha j k}$ are functions of the parameters of the Anderson model satisfying the condition $\mathcal{H}_{\text{tun}} = [\mathcal{H}_0, \mathcal{S}]$

$$w^{(1)}_{\alpha j k} = \frac{\mathcal{V}_{\alpha j k}}{\xi_{\alpha j k} - \varepsilon_j - U} , \ w^{(2)}_{\alpha j k} = \frac{\mathcal{V}_{\alpha j k}}{\xi_{\alpha j k} - \varepsilon_j} . \tag{S14}$$

Now, we expand the Hamiltonian $\mathcal{H}'$ in terms of the operator $\mathcal{S}$ and we restrict ourselves to the leading order given by $\mathcal{H}' \approx \mathcal{H}_0 + (1/2)[\mathcal{S}, \mathcal{H}_{\text{tun}}]$. This approximation is valid for $\Gamma/|\varepsilon_j| \ll 1$ and $\Gamma/|\varepsilon_j + U| \ll 1$. We include Eq. (S13) into $\mathcal{H}'$ and we reduce the operators of the quantum dot subspace into a pseudospin subspace. After taking this approach, the leading term of $\mathcal{H}'$ contains four different Hamiltonians: The first term is a two-electron hopping Hamiltonian $\mathcal{H}_{\text{ch}}$ which takes into account processes of two electrons coming in and out the lead simultaneously. Nevertheless, this term is neglected because we assume that charge conserving processes in the dot dominate over the nonconserving ones. The second term corresponds to an additional dot Hamiltonian $\mathcal{H}'_0$ which vanishes in the pseudospin subspace. The next term is an exchange s-d Hamiltonian which reads

$$\mathcal{H}_{\text{ex}} = \sum_{\alpha \beta i j k q} \mathcal{J}_{\alpha i, \beta j} \hat{S}_l s^l_{ij} C^{\dagger}_{\alpha i k} C_{\beta j q} , \tag{S15}$$

where $\hat{S}_l$ and $s^l_{ij}$ ($l = \{x, y, z\}$) are the pseudospin operator and the coefficients of the Pauli matrices, respectively. $\mathcal{J}_{\alpha i, \beta j}$ is the antiferromagnetic coupling between conduction electrons at different reservoirs defined

$$\mathcal{J}_{\alpha i, \beta j} = \mathcal{V}_{\alpha i k} \mathcal{V}^*_{\beta j q} [g_i(\omega) - g_i(\omega - U) + g_j(\omega) - g_j(\omega - U)], \tag{S16}$$

where $g_i(\omega) = 1/(\omega - \varepsilon_i)$ is the unperturbed retarded Green's function in the Fourier space. The last term is a direct s-d Hamiltonian

$$\mathcal{H}_{\text{dir}} = \sum_{\alpha \beta j k q} \mathcal{K}_{\alpha j, \beta j} C^{\dagger}_{\alpha j k} C_{\beta j q} , \tag{S17}$$

where $\mathcal{K}_{\alpha i, \beta j}$ is the amplitude of the potential scattering term reading

$$\mathcal{K}_{\alpha i, \beta j} = \frac{1}{4} \mathcal{V}_{\alpha i k} \mathcal{V}^*_{\beta j q} [g_i(\omega) + g_i(\omega - U) + g_j(\omega) + g_j(\omega - U)]. \tag{S18}$$

In contrast with Eq. (S15), Eq. (S17) does not depend on the pseudospin meaning that it has no contribution from the pseudospin-flip processes.

The combination of both direct and exchange s-d terms forms the Kondo Hamiltonian defined in the main manuscript.

## VI. PERTURBATION EXPANSION

The aim of this section is to explain the perturbation expansion yielding Eq. (5). This will be performed by expanding the expected value of the current in terms of the amplitudes $\mathcal{K}$ and $\mathcal{J}$.

Following Eqs. (S15) and (S17), the electric current $I_{\alpha j} = -e \frac{dN_{\alpha j}}{dt}$ becomes

$$\hat{I}_{\alpha j} = -\frac{ie}{\hbar} \sum_{\beta k_\alpha k_\beta} \left[ \mathcal{K}_{\beta j, \alpha j} C^{\dagger}_{\beta k_\beta j} C_{\alpha k_\alpha j} + \sum_{l, r} \mathcal{J}_{\beta r, \alpha j} \hat{S}_l s^l_{rj} C^{\dagger}_{\beta k_\beta r} C_{\alpha k_\alpha j} - \text{H. c.} \right] . \tag{S19}$$

We highlight that, in contrast with Sec. I, Eq. (S19) obeys $I_{Lj} + I_{Rj} = \pm \langle dS_z/dt \rangle$ as a charge conservation law. This is due to the fact that Eq. (S19) contains cotunnel processes involving hopping of electrons in both subsystems which change the pseudospin of the double quantum dot. Nevertheless, the sum of all currents is charge-conserved $\sum_{\alpha j} I_{\alpha j} = 0$.

We employ a time-dependent perturbation of the expected value of current $\langle \bar{S}(-\infty, 0) \hat{I}_{\alpha j}(0) \bar{S}(0, -\infty) \rangle$ following the perturbation $\bar{S} = \hat{T} \exp\left[-(i/\hbar) \int dt \mathcal{H}_K(t)\right]$. The leading order of the expansion, following Eq. (S19) and Eqs. (S15) and (S17), reads

$$
\begin{aligned}
I_{\alpha j} = &\frac{2e}{\hbar^2} \sum_{\vec{v}} \left( \tilde{\delta}_1 \mathcal{K}_{\alpha_1 j_1, \beta_1 j_1} \mathcal{K}_{\beta j, \alpha j} + \tilde{x}_1 \mathcal{J}_{\alpha_1 j_1, \beta_1 j_{\beta_1}} \mathcal{J}_{\beta j_{\beta}, \alpha j} \right) \text{Re} \left[ \int_{-\infty}^{0} dt \langle C_{\alpha_1 j_1 k_{\alpha_1}}^{\dagger} C_{\beta_1 j_1 k_{\beta_1}} C_{\beta j_{\beta} k_{\beta}}^{\dagger} C_{\alpha j k_{\alpha}} \rangle \right] \\
&-\frac{2e}{\hbar^2} \sum_{\vec{v}} \left( \tilde{\delta}_1 \mathcal{K}_{\alpha_1 j_1, \beta_1 j_1} \mathcal{K}_{\alpha j, \beta j} + \tilde{x}_2 \mathcal{J}_{\alpha_1 j_1, \beta_1 j_{\beta_1}} \mathcal{J}_{\alpha j, \beta j_{\beta}} \right) \text{Re} \left[ \int_{-\infty}^{0} dt \langle C_{\alpha_1 j_1 k_{\alpha_1}}^{\dagger} C_{\beta_1 j_{\beta_1} k_{\beta_1}} C_{\alpha j k_{\alpha}}^{\dagger} C_{\beta j_{\beta} k_{\beta}} \rangle \right],
\end{aligned} \quad (S20)
$$

where $\vec{v} = \{\alpha_1, k_{\alpha_1}, j_1, \beta_1, k_{\beta_1}, j_{\beta_1}, \beta, k_{\beta}, j_{\beta}, k_{\alpha}\}$, $\tilde{\delta}_1 = \delta_{j_1 j_{\beta_1}} \delta_{j j_{\beta}}$, $\tilde{x}_1 = \sum_{ll'} \langle S_l S_{l'} \rangle s_{j_1 j_{\beta_1}}^{l} s_{j j_{\beta}}^{l'}$ and $\tilde{x}_2 = \langle S_l S_{l'} \rangle s_{j_1 j_{\beta_1}}^{l} s_{j_{\beta} j}^{l'}$. Here, we apply the Wick theorem and replace the expected values with its corresponding lesser and greater Green's functions finding

$$
\begin{aligned}
I_{\alpha j} = &2e \sum_{\beta k_{\beta} j_{\beta} k_{\alpha}} \left( \tilde{\delta}_1 |\mathcal{K}_{\beta j, \alpha j}|^2 + \tilde{x}_1 |\mathcal{J}_{\beta j_{\beta}, \alpha j}|^2 \right) \text{Re} \left[ \int_{-\infty}^{0} dt g_{\beta j_{\beta} k_{\beta}}^{>}(t, 0) g_{\alpha j k_{\alpha}}^{<}(0, t) \right] \\
&-2e \sum_{\beta k_{\beta} j_{\beta} k_{\alpha}} \left( \tilde{\delta}_1 |\mathcal{K}_{\alpha j, \beta j}|^2 + \tilde{x}_2 |\mathcal{J}_{\alpha j, \beta j_{\beta}}| \right) \text{Re} \left[ \int_{-\infty}^{0} dt g_{\alpha j k}^{>}(t, 0) g_{\beta j_{\beta} k_{\beta}}^{<}(0, t) \right].
\end{aligned} \quad (S21)
$$

In order to solve the time integral, we substitute the Green's functions by

$$
g_{\alpha j k}^{<}(t, t') = \frac{i}{\hbar} e^{-\frac{i}{\hbar} \varepsilon_{\alpha j k}(t-t')} f_{\alpha j}(\varepsilon_{\alpha j k}), \quad g_{\alpha j}^{>}(t, t') = -\frac{i}{\hbar} e^{-\frac{i}{\hbar} \varepsilon_{\alpha j k}(t-t')} [1 - f_{\alpha j}(\varepsilon_{\alpha j k})]. \quad (S22)
$$

Additionally, the pseudospin expected value is $\langle S_l S_{l'} \rangle = \delta_{ll'}/4$. Therefore, Eq. (S21) becomes

$$
\begin{aligned}
I_{\alpha j} = &\frac{e\pi}{4\hbar^2} \sum_{\beta} \int d\omega \rho_{\alpha j} \rho_{\beta \bar{j}} |\mathcal{J}_{\beta \bar{j}, \alpha j}|^2 \left[ f_{\alpha j}(\omega) - f_{\beta \bar{j}}(\omega) \right] \\
&+\frac{e\pi}{8\hbar^2} \sum_{\beta} \int d\omega \rho_{\alpha j} \rho_{\beta j} \left( 16 |\mathcal{K}_{\beta j, \alpha j}|^2 + |\mathcal{J}_{\beta j, \alpha j}|^2 \right) \left[ f_{\alpha j}(\omega) - f_{\beta j}(\omega) \right],
\end{aligned} \quad (S23)
$$

where $\rho_{\alpha j}$ denotes the density of states of the fermionic reservoir $\alpha j$. Importantly, in the drag system ($j = 1$), the second term of Eq. (S23) vanishes meaning that only spin-flip processes are taken into account in such approximation. Finally, Eq. (5) of the main text is obtained after replacing $\mathcal{J}_{\beta \bar{j}, \alpha j}$ following Eq. (S16). Remarkably, such expression coincides exactly with the result of a master equation approach assuming cotunneling in the particle-hole symmetry point



\end{document}